\documentclass[runningheads]{llncs}
\usepackage{helvet}
\usepackage{paralist}
\usepackage{xcolor}
\usepackage{graphicx}
\usepackage{comment}
\usepackage{hyperref}
\usepackage{xspace}
\usepackage{amssymb}
\usepackage{amsmath}
\usepackage{enumitem}

\newcommand{\tbif}{{\bf ~~if~~}}
\newcommand{\tbprev}{{\bf prev~}}

\newcommand{\dds}{{DDS}\xspace}
\newcommand{\ddss}{{\dds}s\xspace}

\newcommand{\drts}{DRTS\xspace}

\newcommand{\gencommtype}{\mathfrak{C}}

\newcommand{\cstepa}{\textsc{c-step}}
\newcommand{\s}{\ensuremath{\mathcal{M}}\xspace}
\newcommand{\prog}{\ensuremath{\mathcal{P}}\xspace}

\newcommand{\net}{\ensuremath{\mathcal{N}}\xspace}
\newcommand{\indata}{\ensuremath{D_0}\xspace}

\newcommand{\nodes}{V}
\newcommand{\arcs}{A}

\newcommand{\tup}[1]{\langle #1 \rangle}

\newcommand{\lang}{\textsc{d2c}\xspace}
\newcommand{\schemasym}[1]{\mathcal{#1}}
\newcommand{\ssch}{\schemasym{S}}
\newcommand{\isch}{\schemasym{I}}
\newcommand{\tsch}{\schemasym{T}}

\newcommand{\states}{\Sigma}
\newcommand{\state}{\sigma}
\newcommand{\trel}{\Rightarrow}
\newcommand{\istate}{\state _0}

\newcommand{\const}{\Delta}

\newcommand{\ndb}{\mathit{ndb}}
\newcommand{\channel}{\mathit{nch}}
\newcommand{\anode}{{\tt n}}
\newcommand{\instset}[1]{\mathbb{#1}}
\newcommand{\qinstset}{\instset{C}}
\newcommand{\sinstset}{\instset{S}}
\newcommand{\iinstset}{\instset{I}}

\usepackage{stmaryrd}
\newcommand{\tsfam}[1]{\llbracket #1 \rrbracket}

\newcommand{\stdb}[1]{\mathit{state}(#1)}
\newcommand{\trdb}[1]{\mathit{transp}(#1)}
\newcommand{\trdbtup}[1]{\mathit{transp}\downarrow(#1)}

\newcommand{\trsys}[3]{\ensuremath{\Upsilon_{#1}^{\text{#3}}}}
\begin{document}
\title{Verification of Sometimes Termination of Lazy-Bounded Declarative Distributed Systems\thanks{Published in the online proceedings of the ESSLLI 2021 Student Session.}}
\titlerunning{Sometimes termination of LB-DDSs}
\author{Francesco Di Cosmo\inst{1}}
\authorrunning{F. Di Cosmo}
\institute{Free University of Bozen-Bolzano \email{fdicosmo@unibz.it}}
\maketitle
\begin{abstract}
Declarative Distributed Systems (DDSs) are distributed systems grounded in logic programming. Although DDS model-checking is undecidable in general, we detect decidable cases by tweaking the data-source bounds, the message expressiveness, and the channel-type.
\end{abstract}
\section{Introduction}
Motivated by Declarative Networking \cite{LCGG*09} and Business Process Analysis \cite{BCDDM13}, Declarative Distributed Systems (DDSs) form a model of distributed computing grounded in logic programming. 
A number of similar models were considered in the literature, indicating that the declarative paradigm allows for concise and intuitive implementations of the distributed behavior \cite{NJLS12}. For example, they have been used for
security and provenance in distributed query processing~\cite{ZSAM15,ZMRL*12},
in the analysis of asynchronous event systems~\cite{AAHM05}, and as the core of
the Webdam language for distributed Web applications~\cite{ABGA11}. A common trait of these models is data-centricity, i.e., local node computations consists of queries over relational databases, which provide a close relationship between the programs and their formal semantics. In turn, this feature simplifies the development of analysis tools and techniques \cite{RZWJ*11,MLWRL13}.
In fact, those have been exploited in various settings, but providing only empirical and experimental assessments \cite{RZWJ*11,MLWRL13,CJXL*14,CLJZL15}, or formal models to study distributed query computation strategies \cite{AmNV13,AKNZ14,AGKNS15}, disregarding their temporal evolution.

Our broad research objective is to fill that gap. Specifically we aim at a comprehensive, rigorous study of DDSs verification through model-checking techniques. Unfortunately, the general problem is undecidable \cite{calvanese18}, thus our efforts are directed towards finding constraints to gain decidability, e.g., by placing boundedness conditions on the various system data-sources. However, boundedness itself is semi-decidable and, thus, not satisfactory in verifying DDSs not already known to be bounded.
Our ongoing research tests whether it is possible to remove boundedness conditions in favor of constraints on the message expressiveness. As a case-study, in this paper we contrast sometimes termination of DDSs with propositional messages to reachability of Communicating Finite State Machines (CFSMs) \cite{chambart08}.

In the next Section \ref{sec:prel} we define DDSs and the problem of sometimes termination. Then, in Section \ref{sec:bounded}, we introduce the concept boundedness, delineate our case-study, and sketch the proofs of our results. Finally, in Section \ref{sec:conc} we discuss directions for future works.

\section{Preliminaries}\label{sec:prel}
\subsection{Declarative Distributed Systems}
A DDS is a set of inter-communicating nodes over a network of symmetric channels. In addition, each node has a self-loop channel to communicate with itself. A network comprises only perfect channels, only lossy channels (if messages can be lost during message casting), or only unordered channels (if the message reception order may be different from the sending one). However, for reasons discussed below, in this paper we will not consider lossy channels.

Locally, each node can manipulate only relational databases (DBs). Nodes accept input-DBs, over an input-signature $\isch$, according to one of the following input-policies:
\begin{inparaenum}[(1)]
\item closed, if there is no input (empty input-DB);
\item interactive, if there may be a new input-DB at each computation step;
\item autonomous, if the input is fixed at startup.
\end{inparaenum}
Messages are single relational facts over a transport-signature $\tsch$. They are received one per step according to the asynchronous semantics and their reception triggers node activation. Thus, to allow startup, each self-loop channel is initially populated by the special $\tt start/0$ message. The channel content is an instance of a fixed data structure $\gencommtype_\tsch$ of facts over $\tsch$. Perfect channels are modeled by the $\tt queue_\tsch$ data structure, while unordered channels exploit the $\tt multiset_\tsch$ one. 
Moreover, each node maintains a state-DB over a state-signature $\ssch$, which includes the node name and the neighbor-set, stored in persistent and read-only relations $\tt my\_name/1$ and $\tt my\_neighbor/1$.

Nodes share a common program written in the Datalog-like language D2C \cite{ma16}. In this exposition, we suppose that the reader has some familiarity with stratified Datalog (see \cite{abiteboul95}). Notably, D2C features in-rule operators that allow to \begin{inparaenum}[(1)]
\item refer to the previous state-DB,
\item append addresses to transport literals (to check message reception and trigger message sending) and
\item non-deterministically enforce functional dependencies among the variables in the rule.
\end{inparaenum}
At node activation, a new computation step starts: the node applies its program to its data-sources (input, message, and state), transitions to a new state-DB, and produces and sends in non-deterministic order a set of messages to their respective recipients.

Formally, \lang rules are of the form:
\[
  H \tbif L_1,\ldots,L_n, \tbprev L_{n+1},\ldots,L_m, C
\]
The head $H$ is a state or transport atom. The body literals $L_i$s are transport, state, or input literals. The constraint $C$ is a sequence of inequalities among terms (variables and constants). Transport atoms must always be labeled with a string of the form $@t$, where $t$ is term. As usual in Datalog-like languages, rules must be safe, i.e., each variable should occur at least one time in a positive body literal.
Additionally, among the body literals, a unique choice-predicate in the manner of Saccà and Zaniolo \cite{SaZa90}, of the form $\tt choice(X,Y)$ or $\tt choice(Y)$, can occur, to enforce a non-deterministically chosen functional dependency of the form $X\longrightarrow Y$ or $\varepsilon\longrightarrow Y$, respectively; the latter is a convenient tool to select a single instantiation of $Y$ among many different candidates. Variables occuring in $\tt choice$ predicates must occur in some other standard body-literal.

Intuitively, to evaluate a rule, its variables are instantiated over the node active domain and, if 
\begin{inparaenum}[(1)]
\item the state body literals outside the scope of $\tbprev$ are true in the node state-DB during the current step,
\item the transport body-literals outside the scope of $\tbprev{}$ match the incoming message and their labels identify the sending node,
\item the same applies to body literals inside the scope of $\tbprev$, with the provision that they must refer to the previous computation step, and
\item the constraint sequence is satisfied,
\end{inparaenum}
 then the (now grounded) head is deduced. Deduced transport-heads are sent to the node mentioned in its label (it should match a neighbour ID), while state-heads are added to the current state-DB. Rules are evaluated iteratively until a fix-point is reached.
 Additionally, in case a $\tt choice$ predicate occurs in the rule, the instantiated rules that satisfy the previous conditions are filtered so to enforce a corresponding functional dependency. Technically, the semantics of rules with $\tt choice$ in the manner of Saccà and Zaniolo requires to substitue the rule with three related rules that exploit a recursive cycle with negated atoms, to be interpreted under the stable model semantics. Otherwise, as mentioned above, all explicit negations in the rule body should be stratified, i.e., roughly, the fix-point computation does not contain recursive cycles over negated literals. However, stratification should only apply to state-literals outside the scope of $\tbprev$, since transport atoms and the remaining body literals refers to the fixed incoming message and the state-DB computed during the previous step, which behave as extensional data.
 
 Given a program $P$, an input DB $I$, a previous state DB $S$, and a labeled transport tuple $t@n$, we denote by $\stdb{P,I,S,t@n}$ the new state database computed by the program $P$ over $ S\cup I\cup \{t@n\}$, by $\trdb{P,I,S,t@n,d}$ the set of computed transport relations labeled by $@d$, and by $\trdbtup{P,I,S,t@n,d}$ the set of elements in $\trdb{P,I,S,t@n,d}$ where the label $@d$ has been dropped.

\paragraph{DDS formal semantics.}
With those notions at hand, we can formalize \ddss. Technically, a \emph{\dds} $\s$ is a tuple
$\tup{\net,\isch,\tsch,\ssch,\prog,\indata}$, where:
\begin{itemize}
\item $\net=\tup{\nodes,\arcs}$ is an undirected network graph where each node $v\in\nodes$ has a self-loop;
\item
  $\isch$, $\tsch$, and $\ssch$ respectively denote the input, transport, and state schemas
  of every node in $\s$;
\item $\prog$ is the \lang program run by every node in $\s$;
\item $\indata$ is a local state DB over $\ssch$ representing the initial
  state of each node, and so that it assigns no tuples to $\tt my\_name$ and $\tt
  neighbor$ (these are in fact implicitly set differently for each node, depending on the network topology).
\end{itemize}

While, usually, the semantics of dynamic systems over data are given in terms of relational transition systems (RTSs), for \ddss it is more convenient to exploit \textit{distributed RTSs} (DRTSs), i.e., a distributed version of RTSs that accounts for each node and channel local configuration.
Given a \dds $\tup{\net,\isch,\tsch,\ssch,\prog,\indata}$ with $\net=\tup{\nodes,\arcs}$, and given a data structure $\gencommtype_\tsch$ over $\tsch$, the $\drts$ of the \dds under $\gencommtype_\tsch$ is a tuple $\tup{\net,\const,\ssch,\tsch,\states,\istate,\ndb,\channel,{\trel}}$, where:
\begin{itemize}
\item $\states$ is a (possibly infinite) set of states;
\item $\const$ is a (possibly infinite) countable domain of constants;
\item $\istate \in \states$ is the initial state;
\item $\ndb$ is a function that, given a state $\state \in \states$ and a node $\anode \in \nodes$, returns a
  corresponding DB instance of $\ssch$ over $\Delta$ for $\anode$;
\item $\channel$ is a function that, given a state $\state \in \states$ and two nodes $\anode_1,\anode_2 \in \nodes$ such that $\tup{\anode_1,\anode_2} \in \arcs$, returns an instance of $\gencommtype_\tsch$ storing the pending messages from $\anode_1$ to $\anode_2$;
\item ${\trel} \subseteq \states\times\states$ is a transition relation between
  states.
\end{itemize}

To define the execution semantics under the interactive input-policy, we introduce the following relation $\cstepa$.
\[\cstepa \subseteq \prod_{n\in V}\sinstset  \times \prod_{a\in A}\qinstset \times \arcs
\times \iinstset \times \prod_{n\in V}\sinstset  \times \prod_{a\in A}\qinstset
\]
where $\sinstset$ is the set of all possible state DB of $\ssch$, $\iinstset$ is the set of all possible state DB of $\isch$ and $\qinstset$ is the of all possible instantiations of $\gencommtype_\tsch$, always over $\Delta$.
Specifically, given $S, S'\in \prod_{n\in V}\sinstset$, and $C, C'\in \prod_{n\in V}\qinstset$, a channel ${\tt (s,d)} \in \arcs$,
and an input database $I\in \iinstset$,
we have that $\tup{S, C, ({\tt s},{\tt d}),I,S',C'} \in \cstepa$ if and only if there exists a message tuple
$t\in C_{({\tt s},{\tt d})}$ such that:
\[
S'_n=
\begin{cases}
\stdb{P,I,S,t@s} & \text{if }n=d\\
S_n & \text{otherwise}
\end{cases}
\]
\[
C'_{(n,m)}=
\begin{cases}
C_{(s,d)}\setminus\{t\} & \text{if }n=s\text{ and }m=d\\
C_{(d,m)}\cup \trdbtup{P,I,S_n,t@d,m}& \text{if }n=d\\
C_{(n,m)} & \text{otherwise}
\end{cases}
\]

Finally, we define the \emph{interactive transition system} of $\s$,
written $\trsys{\s}{a}{int}$,
as the \drts
$\tup{\net,\const,\ssch,\tsch,\states,\istate,\ndb,\channel,{\trel}}$, where:
\begin{itemize}
\item $\Sigma\subset \prod_{n\in \nodes}\sinstset  \times \prod_{a\in \arcs}\qinstset$ and, for each $\state \in \Sigma$ of the form $\state=((S_n)_{n\in \nodes},((C_c)_{c\in\arcs}))$, $\ndb(\state,n)=S_n$ and $\channel(\sigma,s,d)=C_{(s,d)}$;
\item $\istate=((\indata)_{n\in\nodes},(C_{c})_{c\in\arcs})$, where $C_{(s,d)}=\emptyset$ if $s\neq d$ and $C(s,d)=\{{\tt start}\}$ otherwise, where ${\tt start}$ is  a special 0-ary transport tuple;
\item The extensions of $\states$ and ${\trel}$ are defined by simultaneous induction as follows:
  \begin{enumerate}
  \item$\istate\in\Sigma$;
  \item if $(S,C) \in \Sigma$, then, for each $ \tup{S,C, ({\tt s},{\tt d}),I,S',C'}\in \cstepa$, it is true that $(S',C')\in \Sigma$ and $(S,C) \trel (S',C')$.
  \item if $(S,C)\in \Sigma$ and, for each $c\in\arcs$, $C_c=\emptyset$, then $(S,C)\trel (S,C)$.
  \end{enumerate}
 \end{itemize}

In the case of the autonomous input-policy, the
semantic presented above needs to be modified so as to take into
account that the input is provided only at the beginning of the
computation and consequently stays rigid over time. Clearly, each
computation may be associated to a different initial input. We can then imagine
that
the execution semantics is given, in this case, by an infinite set $\tsfam{\trsys{\s}{a}{aut}}$ of transition
systems, each associated to a different global input DB $I\in \iinstset$. For each input DB $I$, the associated transition system is defined as above, with the provision that the relation $\cstepa$ is modified into the relation $\cstepa^I$, where the component $\iinstset$ is dropped in favor of the fixed parameter $I$.
Since closed dds can be seen as autonomous ddss over an empty input schema, the closed execution semantics is defined by the RTS $\trsys{\s}{a}{clo}$ induced by $\cstepa^\emptyset$.

\paragraph{Sometimes termination}
We say that a \dds \textit{sometimes terminates} if its \drts exhibits at least one run which reaches a configuration where no node-activation can be triggered anymore, i.e., all channels are empty. Over lossy networks it is a trivial problem, since to lose all messages implies termination; that's why we do not consider lossy channels. However, it is undecidable over perfect and unordered networks \cite{calvanese18}, irrespective of the input policy.

\begin{example}
In this toy-example we present a single-node closed-input \dds that models a game: a player moves a pebble in a maze attempting to reach an exit. That \dds is designed so that it sometimes terminates if and only if the player as at least one possibility to win. For this example, the channel type is irrelevant.

The initial state-DB contains a description of a maze with a finite number of positions, each one of them has at most four outgoing paths named $\tt up$, $\tt down$, $\tt right$, and $\tt left$. The available paths at each position are stored in the relation $\tt path/3$, e.g., $\tt path(a,b,up)$ means that at position $\tt a$ there is a path to position $\tt b$ going $\tt up$. The following rules make those information persistent throughout the whole \dds computation:
\[
    \begin{array}{@{}r@{}l@{}}
      \tt path(X,Y,Z) ~\tbif~ & \tt \tbprev path(X,Y,Z).
    \end{array}
\]
The player position is stored in the predicate $\tt player/1$. Since it will change in time, it does not need to be persistent. Instead, that is true for the position of the exit stored in $\tt exit/1$:
\[
    \begin{array}{@{}r@{}l@{}}
      \tt exit(X) ~\tbif~ & \tt \tbprev exit(X).
    \end{array}
\]
The player choices are modeled through D2C non-determinism. At each computation step, the node non-deterministically selects an available path and makes a step in that direction:
\[
    \begin{array}{@{}r@{}l@{}}
      \tt player(X) ~\tbif~ & \tt path(p,X,D), choice(D)  ~\tbprev player(p).
    \end{array}
\]
To encode the winning condition, a $\tt win/0$ persistent propositional flag is deduced when the player and exit positions coincide:
\[
    \begin{array}{@{}r@{}l@{}}
      \tt win ~\tbif~ & \tt exit(X), player(X).\\
      \tt win ~\tbif~ & \tt \tbprev win.
    \end{array}
\]
To perform steps, the node continuously sends to himself a $\tt wakeUp/0$ propositional message, unless the winning condition has been met, i.e., the $\tt win$ flag is in the state-DB:
\[
    \begin{array}{@{}r@{}l@{}}
      \tt wakeUp@X ~\tbif~ & \tt my\_name(X) ~\tbprev \neg win.
    \end{array}
\]

Thus, the self-loop channel gets empty along a \dds run if and only if in that run a configuration where the player and exit positions coincide is reached, i.e., there is at least one sequence of choices that makes the player win.
\end{example}

\section{Lazy-bounded DDSs and CFSMs}\label{sec:bounded}
To get decidable cases of DDS verification, it is sufficient to require \textit{boundedness} over channel sizes and source-DBs (input- or state-DB), i.e., there is a natural number $b$ such that, along all DDS runs, at each computation step, the DB mentions at most $b$ constants. However, an infinite amount of data can still flow through the DDS along infinite runs. The following combinations of boundedness conditions and input-policies yield decidable verification of sophisticated languages (see \cite{calvanese18}) that mix first order logic, to express queries over the nodes and the channel local configurations, with computation tree logic, to analyze the DDS temporal behavior:
\begin{inparaenum}[(1)]
\item channel-bounded closed-input DDSs;
\item channel- and state-bounded interactive-input DDSs;
\item channel-, state-, and input-bounded autonomous-input DDSs.
\end{inparaenum}
Moreover, those combinations are essential, in the sense that no boundedness condition can be dropped without losing decidability. While interactive DDSs may seem harder to verify wrt autonomous ones, they actually require less boundedness conditions to enjoy of decidable verification. That can be motivated by the fact that a changing input decreases computational power. To explain that with an analogy, imagine a Turing machine in which a part of the tape 
changes uncontrollably, allowing the reading of a tape cell only if the machine head is in the right place at the right time, without the possibility of performing a persistent writing.

Unfortunately, while the bound $b$ can be computed, if it exists, in PSPACE in the DDS initial state complexity, it is undecidable to check its existence \cite{BCDDM13}. Thus, we are looking for alternative conditions that retain decidability but are also themselves decidable. We start by lifting channel-boundedness, since it applies to all input-policies, and consider \textit{Lazy-Bounded DDSs} (LB-DDSs), i.e.,
\begin{inparaenum}[(1)]
\item closed-input DDSs,
\item state-bounded interactive-input DDSs, and
\item state- and input-bounded autonomous-input DDSs.
\end{inparaenum}
Furthermore, instead of bounding the channels, we constrain message expressiveness by bounding the arities in the transport-signature. As a preliminary case, we consider \textit{propositional LB-DDSs} (PLB-DDSs), i.e., LB-DDSs with propositional transport-signature.

Intuitively, lazy-boundedness amounts to finite control and propositional transport-signatures amount to finite message-sets. Thus, PLB-DDS sometimes termination seems very similar to reachability of CFSMs, i.e., distributed systems where nodes are finite state machines with reading and writing transitions. That problem was extensively studied, even parameterized over channel type \cite{chambart08}. In fact, CFSM reachability (already over target configuration-sets with empty channels) is undecidable over perfect networks (already over single-node networks) but decidable over lossy and unordered networks \cite{chambart08}. Unfortunately, we cannot directly transfer those results to PLB-DDSs because of the following issues:
\begin{inparaenum} [(1)]
\item despite boundedness, PLB-DDSs may still have an infinite number of configurations;\label{infinite-RTS}
\item PLB-DDSs receive and send messages in one step, while CFSMs decouple reading and writing transitions;\label{steps-transitions}
\item sometimes termination is a sub-case of full-fledged reachability;\label{termination-reachability}
\item the PLB-DDS activation-policy does not apply to CFSMs.\label{activation}
\end{inparaenum}
However, we sketch reductions that solve those issues, from PLB-DDSs sometimes termination to CFSMs reachability, over unordered networks, and vice versa, over perfect networks.

\begin{example}
The \dds in the previous example exploits the closed input-policy and, thus, it is trivially state-bounded. As a matter of facts, since at each step a message is consumed and at most one is produced, the self-loop channel always contains at most one message, i.e., it is 1-channel-bounded. However, since the transport signature is propositional, we can add complexity to the game and still model it with a PLB-DDS where channel-boundedness is lifted.

Suppose that, at each turn, at most one available path is blocked. To make the blocking behavior dependant on the player movements, each position is annotated with a set of path names. When the player moves to a position, he writes each annotated path name on individual cards and put them into a deck. Before moving again, he randomly draw a card from the deck and move through an available path different from that mentioned in the drew card. Thus, the probability that a specific path will be blocked depends on the maze design and the player previous movements. Additionally, the deck always contains a single, special card that does not block any path.

The position annotation is formalized by the persistent predicate $\tt card/2$, e.g., $\tt card(p,up)$ forces the player entering in position $p$ to put in the deck a card annotated with $\tt up$. Since the deck can contain an unbounded number of copies of each card, it cannot be encoded inside the state-DB, which is a set. Thus, we have to pt for unordered channels, so that the deck can be encoded in the self-loop channel, which happens to be a multi-set. Also, to store the card on the propositional channel we need to convert the path names, which are terms, into propositional flags. The following rules models the insertion of cards into the channel:
\[
\begin{array}{@{}r@{}l@{}}
  \tt collect(Y) ~\tbif~ & \tt player(X), card(X,Y).\\
  \tt up@X ~\tbif~ & \tt my\_name(X), collect(up).\\
  \tt down@X ~\tbif~ & \tt my\_name(X), collect(down).\\
  \tt left@X ~\tbif~ & \tt my\_name(X), collect(left).\\
  \tt right@X ~\tbif~ & \tt my\_name(X), collect(right).\\
\end{array}
\]
The following rules put the special card $\tt none/0$ into the channel at system startup and each time it is received, thus making sure that the channel contains always exactly one $\tt none$ card:
\[
\begin{array}{@{}r@{}l@{}}
  \tt none ~\tbif~ & \tt my\_name(X), start@X.\\
  \tt none@X ~\tbif~ & \tt my\_name(X), none@X.
\end{array}
\]
To activate those rules only when the player has not yet won, they should be extended with a $\tt \neg \tt win$ body-literal.

Those rules work on top of those of the previous example, with the provision that the one mentioning $\tt wakeUp$ must be dropped, since activation is enforced by the reception of cards.
\end{example}

\paragraph{Decidability over unordered networks.}
Only issues \ref{infinite-RTS} and \ref{steps-transitions} appear to be relevant in simulating PLB-DDSs via CFSMs, over the same unordered network. To mitigate issue \ref{infinite-RTS}, we can exploit the fact that D2C, as virtually all Datalog-like languages, enforce uniformity, i.e., from isomorphic state-DBs it is possible to transition only to isomorphic ones. Then, by knowing the bound $b$, we can fix a finite set of constants to use as active domain and produce a finite number, exponential in the DDS initial configuration, of states over that domain. It can be proved that those are enough to represent all PLB-DDS local configurations and transitions, up to isomorphisms \cite{belardinelli17}; hence, we can use them as the CFSM node state-set.

To solve issue \ref{steps-transitions}, we pair each PLB-DDS message reception with a CFSM reading transition that triggers a sequence of writing transitions. That sequence can be effectively constructed by applying the DDS program to the state and read message mentioned in the reading transition, but it calls for the introduction of new auxiliary states to avoid the occurrence of unwanted transitions during the writing sequence.

It now suffices to prove that, since the CFSM and PLB-DDS networks share the same network type and topology, the CFSM can reach a configuration where all channels are empty and no node is in an auxiliary state if and only if the PLB-DDS sometimes terminates. Thus, even if the size of the built CFSM is not polynomial with respect to the PLB-DDS initial configuration complexity, we can still exploit CFSM reachability (decidable over unordered networks) to check PLB-DDS sometimes termination. Thus, PLB-DDS over unordered channels is decidable.

\paragraph{Undecidability over perfect networks.}
Over perfect networks, we transform single-node CFSM reachability of target configurations with empty channels to closed-input PLB-DDS sometimes termination.
D2C is powerful enough to encode the local CFSM node behavior employing a single PLB-DDS node: the initial state and the transition-set can be provided in the initial state-DB, while the transition semantics and message passing can be enforced by a couple of D2C rules. Thus, we transform the CFSM node in its PLB-DDS counterpart, called \textit{simulator}, and provide to it the extension of the reachability target-set. Hence, the simulator can always check whether its state is a target, in which case it stops.

Only issues \ref{termination-reachability} and \ref{activation} are relevant. To solve them, we attach to the simulator two children, called \textit{activator} and \textit{terminator}. 
The activator is responsible for producing a non-deterministic number of activation messages, exploiting non-deterministic functional dependencies, and sending them to the simulator and activate it to (possibly) simulate writing transitions even when its self-loop channel is empty. To do so, boolean values are stored in the persistent relation $\tt bool/1$ and the deduction of a persistent $\tt stopCreation/0$ flag signals that no more activation messages should be produced:
\[
\begin{array}{@{}r@{}l@{}}
  \tt create(Y) ~\tbif~ & \tt bool(Y), choice(Y) ~\tbprev \neg stopCreation.\\
  \tt stopCreation ~\tbif~ & \tt create(false).\\
  \tt wakeUp@X ~\tbif~ & \tt my\_name(X), create(true).\\
  \tt wakeUp@Y ~\tbif~ & \tt wakeUp@X, my\_name(X), my\_neighbor(Y), create(true), X \neq Y.
\end{array}
\]
The activator needs to sends $\tt wakeUp$ messages also to himself to make sure that it activates until the production mechanism stops.

The terminator continuously writes messages to its self-loop channel, avoiding termination, unless a $\tt stop$ message is received by the simulator upon a  successful reachability check, in which case the slave stops, signaled by the flag $\tt stopped/0$, and consumes its self-loop content without reacting:
\[
\begin{array}{@{}r@{}l@{}}
  \tt wakeUp(X) ~\tbif~ & \tt my\_name(X) ~\tbprev \neg stopped.\\
  \tt stopped ~\tbif~ & \tt stop@X, my\_neighbor(X).
\end{array}
\]
Now, it must be proved that the single-node CFSM exhibits at least one run that reaches a target if and only if there is at least one PLB-DDS run in which the simulator locally detects reachability, if and only if the terminator stops, if and only if the PLB-DDS sometimes terminates.
Thus, if PLB-DDS sometimes termination over perfect networks were decidable, we could use it to check single-node CFSM reachability of empty-channel configurations, which is undecidable. Thus, PLB-DDS over perfect channels is undecidable.

\section{Conclusions}\label{sec:conc}
We have introduced PLB-DDSs and sketched reductions to and from CFSM reachability to chart the decidability of PLB-DDS sometimes termination parameterized with respect to the channel type. Those indicate that PLB-DDS sometimes termination is undecidable over perfect networks and decidable over unordered ones. We argue that similar results apply to sometimes termination of unary-transport LB-DDS. To that end, we plan to build similar reductions to and from reachability of DATA-CFSM \cite{aiswarya20}, where messages range over an infinite data-set. Additionally, to make the case over lossy channels non-trivial, we plan to place probabilities on the channel imperfections and the termination property. That would require the exploitation of probabilistic model-checking techniques. Finally, to exploit decidability through boundedness conditions, i.e., without lifting them, we are investigating D2C syntactic fragments that capture boundedness, i.e., they are guaranteed to be bounded and each bounded DDS can be specified in that fragment. Nevertheless, in that case the problem of D2C program equivalence to one in that fragment would be necessarily undecidable.

\bibliographystyle{splncs04}
\bibliography{string-tiny.bib,biblio.bib,krdb.bib}
\end{document}